\documentclass[10pt,letterpaper,sort&compress]{article}
\usepackage[top=0.85in,left=2.75in,footskip=0.75in]{geometry}

\usepackage{amssymb}
\usepackage{amsmath}
\usepackage{float} 
\usepackage{hyperref}
\usepackage{subfigure}
\usepackage{url}
\usepackage{graphicx}
\usepackage{amsmath}
\usepackage{float}
\usepackage[right]{lineno}
\usepackage[bigfiles]{pdfbase}
\usepackage{longtable}
\usepackage{booktabs, tabularx}
\usepackage{adjustbox}
\usepackage{multimedia}
\usepackage[sort&compress,numbers]{natbib}
\usepackage{xcolor}
\usepackage{epstopdf}

\ExplSyntaxOn
\NewDocumentCommand\embedvideo{smm}{
  \group_begin:
  \leavevmode
  \tl_if_exist:cTF{file_\file_mdfive_hash:n{#3}}{
    \tl_set_eq:Nc\video{file_\file_mdfive_hash:n{#3}}
  }{
    \IfFileExists{#3}{}{\GenericError{}{File~`#3'~not~found}{}{}}
    \pbs_pdfobj:nnn{}{fstream}{{}{#3}}
    \pbs_pdfobj:nnn{}{dict}{
      /Type/Filespec/F~(#3)/UF~(#3)
      /EF~<</F~\pbs_pdflastobj:>>
    }
    \tl_set:Nx\video{\pbs_pdflastobj:}
    \tl_gset_eq:cN{file_\file_mdfive_hash:n{#3}}\video
  }
  \pbs_pdfobj:nnn{}{dict}{
    /Type/RichMediaInstance/Subtype/Video
    /Asset~\video
    /Params~<</FlashVars (
      source=#3&
      skin=SkinOverAllNoFullNoCaption.swf&
      skinAutoHide=true&
      skinBackgroundColor=0x5F5F5F&
      skinBackgroundAlpha=0.75
    )>>
  }
  \pbs_pdfobj:nnn{}{dict}{
    /Type/RichMediaConfiguration/Subtype/Video
    /Instances~[\pbs_pdflastobj:]
  }
  \pbs_pdfobj:nnn{}{dict}{
    /Type/RichMediaContent
    /Assets~<<
      /Names~[(#3)~\video]
    >>
    /Configurations~[\pbs_pdflastobj:]
  }
  \tl_set:Nx\rmcontent{\pbs_pdflastobj:}
  \pbs_pdfobj:nnn{}{dict}{
    /Activation~<<
      /Condition/\IfBooleanTF{#1}{PV}{XA}
      /Presentation~<</Style/Embedded>>
    >>
    /Deactivation~<</Condition/PI>>
  }
  \hbox_set:Nn\l_tmpa_box{#2}
  \tl_set:Nx\l_box_wd_tl{\dim_use:N\box_wd:N\l_tmpa_box}
  \tl_set:Nx\l_box_ht_tl{\dim_use:N\box_ht:N\l_tmpa_box}
  \tl_set:Nx\l_box_dp_tl{\dim_use:N\box_dp:N\l_tmpa_box}
  \pbs_pdfxform:nnnnn{1}{1}{}{}{\l_tmpa_box}
  \pbs_pdfannot:nnnn{\l_box_wd_tl}{\l_box_ht_tl}{\l_box_dp_tl}{
    /Subtype/RichMedia
    /BS~<</W~0/S/S>>
    /Contents~(embedded~video~file:#3)
    /NM~(rma:#3)
    /AP~<</N~\pbs_pdflastxform:>>
    /RichMediaSettings~\pbs_pdflastobj:
    /RichMediaContent~\rmcontent
  }
  \phantom{#2}
  \group_end:
}
\ExplSyntaxOff

\def\be{\begin{equation}}
\def\lan{\left\langle}
\def\ran{\right\rangle}
\def\ee{\end{equation}}
\def\barr{\begin{array}}
\def\earr{\end{array}}
\def\l{\left}
\def\r{\right}
\def\dis{\displaystyle}
\def\ed{\end{document}}

\newlength\savedwidth

\raggedright
\usepackage[aboveskip=1pt,labelfont=bf,labelsep=period,justification=raggedright,singlelinecheck=off]{caption}

\makeatletter
\renewcommand{\@biblabel}[1]{\quad#1.}
\makeatother

\usepackage{lastpage,fancyhdr,graphicx}
\usepackage{epstopdf}
\fancyheadoffset[L]{2.25in}
\fancyfootoffset[L]{2.25in}
\begin{document}
\vspace*{0.2in}

\begin{flushleft}

{\Large
\textbf\newline{COVID anomaly in the correlation analysis of S\&P 500 market states}
}
\newline
\\
M. Mija{\'i}l Mart{\'i}nez-Ramos\textsuperscript{1\P},
Manan Vyas\textsuperscript{1\P*},
Parisa Majari\textsuperscript{1\P},
Thomas H.  Seligman\textsuperscript{1, 2\P}
\\
\bigskip
\textbf{1} Instituto de Ciencias Físicas - Universidad Nacional Autónoma de México,  Cuernavaca, 62210,  Morelos,  México
\\
\textbf{2} Centro Internacional de Ciencias AC - UNAM, Avenida Universidad 1001,  UAEM,  Cuernavaca, 62210,  Morelos,  México
\\
\bigskip

* manan@icf.unam.mx (MV)

\P These authors contributed equally to this work

\end{flushleft}

\section*{Abstract}

Analyzing market states of the S\&P 500 components on a time horizon January 3, 2006 to August 10,  2023, we found the appearance of a new market state not previously seen and we shall discuss its possible implications as an isolated state or as a beginning of a new general market condition.  We study this in terms of the Pearson correlation matrix and relative correlation with respect to the S\&P 500 index.  In both cases the anomaly shows strongly.

\section*{Introduction}

Market states, introduced in 2012 \cite{SciRep2012} on the basis of the correlations of returns, have seen numerous applications in financial market studies and also beyond  \cite{ Zhou_2018, PhysRevE.97.052312, Tang_2018, NJP2018, Springer2019, arXiv2020, Wang_2020,Nie_2020,Heckens_2020, JAMES-2022,  JSM2022,PhyA2022}.  As far as financial markets are concerned there is some additional information obtained beyond the ``State of the Market'' associated to the largest eigenvalue of the correlation matrix of returns \cite{MSBook}; this shows some structure in the Market and differences between markets, that might well relate to their efficiency.  Also risk assessment in situations associated to crashes are of some interest \cite{NJP2018,  arXiv2020}. Yet these studies indicate that the largest eigenvalue or equivalently the average correlation,  which is very strongly correlated with the largest eigenvalue \cite{MSBook,  PRL1999}, seem to dominate the picture.  Dimensionally scaled dynamics have been shown \cite{NJP2018, Springer2019, Heckens_2020, arXiv2020, JSM2022,PhyA2022} and they seem to confirm this idea.  The time trajectory in the space of correlation matrices \cite{Chap23} visits the clusters over longer time horizons. 

Recently some attention was drawn to the use of projected correlations \cite{Heckens_2020, JSM2022, PhyA2022} eliminating the largest eigenvalue which in turn are compared \cite{RC2023} to the use of relative correlations \cite{KenStu, ANZJS2004}.  To some extent this seems to be due to the fact that the number of independent matrix elements is too large to produce a clear picture.  A fruitful idea may be to use pattern recognition {  techniques to visualize these systems}.  Considering that we have  $N = 322$ S\&P 500 stocks, in a time horizon January 3, 2006 to August 10, 2023, it would lead to $N(N+1)/2 =  52,003$ variables.  An analysis for market sectors rather than stocks successfully reduces this \cite{SecAna2023} using coarse grained correlation matrices introduced in  \cite{Heckens_2020, JSM2022, PhyA2022} to symmetric sectorial matrices. { Even these matrices based on ten sectors will produce 55 variables. } Further reductions via coarse grainings seem feasible and also show some success \cite{CoGr2023}. Yet in the last case taken to its extreme of $2 \times 2$ matrices with but three parameters seems interesting \cite{CoGr2023} but not entirely satisfying, in particular because of an arbitrariness that arises.  In previous work it proved useful to use the power map \cite{Guhr_2003, Guhr_2010, THS_2013} to reduce fluctuations. As we wish to eliminate the effect of the average correlation to emphasize subtler correlations, we shall first proceed without this tool. 

In this scenario we found in a new analysis with a time horizon January 3, 2006 to August 10, 2023 that includes the COVID-19 pandemic  \cite{WHO} as well as its aftermath as far as it is known.  We shall find a remarkable fact, namely that in 2020 an entirely new state appears that has not appeared in the time interval January 3, 2006 to December 31, 2019.  {  We will call it the ``COVID state''.} For a period of several months it entirely dominates the picture and then seems to taper off.  Among the intriguing features that do appear, is the fact that the {  COVID state} does not appear in March 2020 \cite{WHO} but about three months later {  on {1st June 2020  \footnote{The exact date depends slightly on the number of states, time horizon and size of the epochs used.}}}. This feature at least can already be well understood,  because that period corresponds to a crash, which is well identified with panic sell-off and for which the average correlation is the main factor. This induces us to use the concept of relative correlation  \cite{KenStu, ANZJS2004} and as reference time series,  we use the simplest time series, namely the S\&P 500 index itself.  We also look into the behavior of relative correlations discussed in detail in Ref. \cite{RC2023}.  Indeed this relative correlation matrix will now have the COVID state as its dominant component with the largest average relative correlation and with this measure it will start in March 2020.   This  establishes the appearance of a new market state. What we can not determine is, if the {  COVID} state indicates the beginning of a changed market behavior or if it essentially ends with COVID.

{  In the next section,  we describe the data and techniques we use.  The following section gives numerical results for state evolution, transition matrices, {  distributions of correlation matrix elements over the total time horizon. and participation ratios}.  Finally, conclusions and an outlook are presented. We also show some illustrations of clustering of correlation matrices in the supplemental material \cite{SM}.}

\renewcommand{\thefootnote}{\arabic{footnote}}

\section*{Data and techniques used}
\label{sec1}

We choose the stocks of S\&P 500 index as they represent the most important quoted companies of the US market \cite{yf}.  From these stocks, we select all those that within the time horizon January 3, 2006 to August 10, 2023 have no more than two consecutive trading days without a quote  ($T = 4431$ total trading days). The number of stocks is thus reduced to 322 and the corresponding stocks are listed in the supplemental material \cite{SM}. 

For the US-Market, as represented by the stocks making up the S\&P 500 index, we find that market states are roughly ordered according to their average correlation as long as we don't choose too large a time scale for epochs.  We divide the total time horizon $T$ into epochs of 20 trading days and use logarithmic returns $r$ between these days as the dataset,  given the adjusted closing price $p_i(t)$ of trading day $t$ for stock $i$,
\be
r_i(t) = \log \left[ \displaystyle\frac{p_i(t)}{p_i(t-1)} \right] \;.
\label{eq:1}
\ee
For the corresponding returns, we assume zero for the days without closing quote while the return for the active trading day is computed using last active trading day.  Using these returns time series, we calculate Pearson correlation matrix\footnote{For the results presented in the paper, we use the formula for the Pearson correlation matrix elements although, at least for some of the epochs, time series are not even weakly stationary.} $C$ with matrix elements given by \cite{KenStu, MSBook}
\be
C_{i,j} = \dis\frac{\lan r_i \; r_j \ran - \lan r_i \ran \lan r_j \ran}{\sigma_i \, \sigma_j} \;,
\label{eq:2}
\ee
with $\sigma$ is the standard deviation of the respective return time series for the stocks.

The relative correlation $RC$ between two return time series $r_i$ and $r_j$ with respect to the S\&P 500 index returns time series $r_{SP}$ is defined as,
\be
RC_{ij:SP} = \dis\frac{C_{i,j}-C_{i,SP} \, C_{j,SP}}{\l[ \l(1-C_{i,SP}^2\r) \l(1-C_{j,SP}^2\r)\r]^{1/2}} \;.
\label{eq:3}
\ee
Here, $C_{i,j}$ are the Pearson correlation coefficients defined by Eq. \eqref{eq:2}.

Using the S\&P 500 data from January 3rd 2006 to August 10th 2023, we divide the total time horizon ($T = 4431$ total trading days) in epochs of 20 trading days with one day shift and we shall analyze the time evolution of market states by clustering 322 $\times$ 322 dimensional correlation matrices of the 322 stocks that were quoted throughout the time horizon with interruption no longer than two consecutive trading days. For the corresponding returns, we assume zero for the days without closing quote, while the return for the active trading day is computed using last active trading day.  We use the $k$-means clustering formalism following the lines of \cite{NJP2018, RC2023} and show the case of five and six market states which seem appropriate for the S\&P 500 data; other state numbers will be discussed in the supplemental material \cite{SM}.  {  We have verified the robustness of COVID state using the $k$-means clustering by increasing the number of clusters from 5 to 12,  both for the Pearson and relative correlations.  Also, we assign the average correlation of the cluster as the specific property of the state.}
 
\section*{Results and discussion}
\label{sec2}

\subsection*{Time evolution of market states}

\begin{figure}[!h]
         \includegraphics[width=13.5cm]{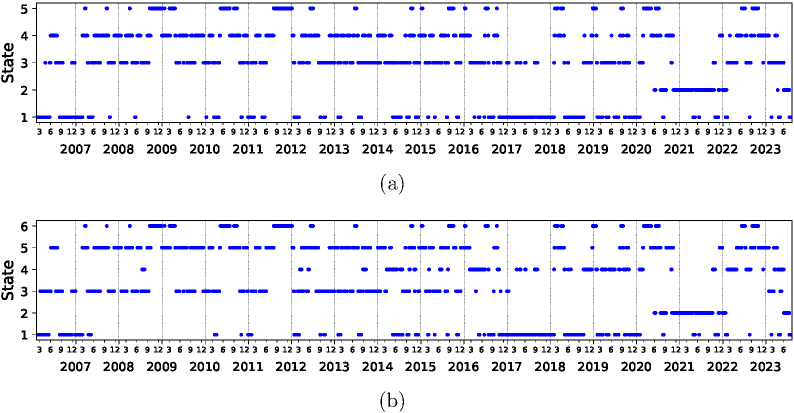}
\caption{Time evolution of market states of the S\&P 500 data using Pearson correlation matrix $C$ defined by Eq. \eqref{eq:2} in a time horizon from January 3rd 2006 to August 10th 2023 with an epoch of 20 trading days. Pearson correlation matrix elements are computed using logarithmic return time series of adjusted closing prices.  Frame (a) and (b) show the cases of five and six states, respectively. The market states are arranged in order of increasing average correlations.  The average correlations for the states are (a) 0.17, 0.27, 0.30, 0.44, 0.61 and (b) 0.16, 0.26, 0.28, 0.31, 0.44, 0.61, respectively.}
\label{fig:1}
\end{figure}

The principal result of this paper is seen in Figure \ref{fig:1} where the time sequence of the five and six states is displayed.  Indeed, these figures show one very notable feature.  State 2 does not appear before June 1st 2020 and then almost uninterruptedly dominates the situation until February 1st 2022 where it peters out.  As the clusters are numbered according to the average correlation it is clear that state 2 corresponds to fairly low average correlation of $\sim 0.26$  but is separated from other low average correlation states.  For comparison, we also show the market evolution for time period between January 3rd 2006 to December 31st 2019 in supplemental material \cite{SM}. Also shown in supplemental material \cite{SM} is the 3D view of the correlation matrices after subjecting them to dimensional scaling according to the recipe given in \cite{borg2005modern}. 

\begin{figure}
           \includegraphics[width=13.5cm]{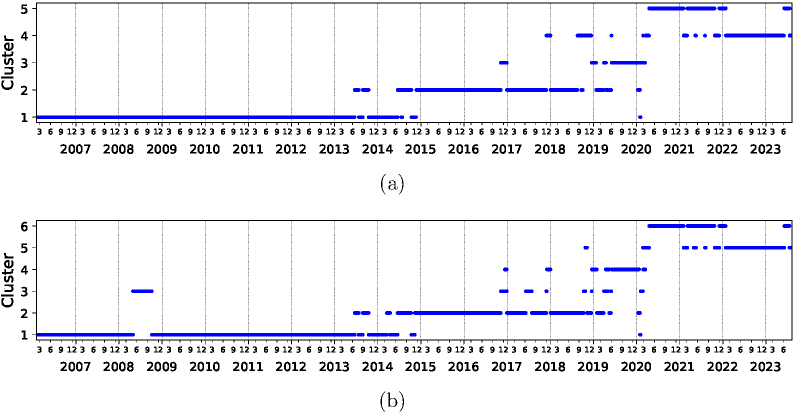}
\caption{Clustering image of the evolution of the relative correlations $RC$  with respect to S\&P 500 index defined by Eq. \eqref{eq:3} with time horizon and epoch length as in Fig. \ref{fig:1}.  The clusters are arranged in order of increasing average relative correlations.  The average relative correlations for the clusters are (a) 0.014, 0.015, 0.019, 0.042, 0.083 and (b) 0.013, 0.015, 0.018, 0.024, 0.048, 0.084, respectively.  Note that the cluster 5 in frame (a) and cluster (6) in frame (b) start approximately three months earlier than the start date of state 2 in  Fig.  \ref{fig:1}. }
\label{fig:R}
\end{figure}

COVID started in March 2020 but the corresponding state 2 of Pearson correlation $C$,  defined in Eq. \eqref{eq:2}, appears only in June 2020. There is a simple explanation for this as the initial financial panic of COVID ended in June 2020. This behavior is associated to the largest eigenvalue and we expect the S\&P 500 index to reflect that.  We therefore look at the evolution of market using relative correlations $RC$, defined in Eq. \eqref{eq:3},  and show the results in Fig.  \ref{fig:R}.  We see that the COVID state is rather isolated but now begins in March 2020. Otherwise, the properties of this state are rather similar - this state will begin in March 2020 and peter out at the approximately same time as with $C$.  The main difference between the corresponding states is the different starting date.  Indeed this state shows the highest average relative correlation and therefore, may also be an important tool to find additional relevant variables { } for the market,  besides the highest eigenvalue of the Pearson correlation matrix.  Other techniques to identify the subtler correlations are discussed in detail in \cite{RC2023}. 

\subsection*{Transition matrices}

\begin{figure}[ht]
          \includegraphics[width=13.5cm]{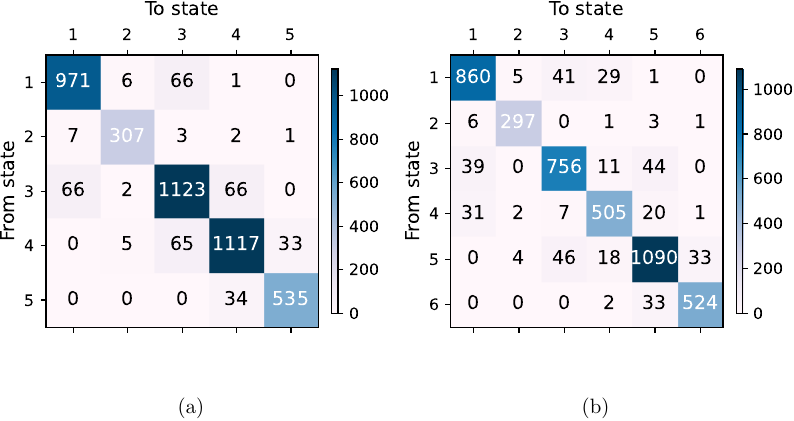}
\caption{Transition matrices showing jumps between different market states shown in Fig. \ref{fig:1} with (a) five clusters and (b) six clusters.  The transition matrices are nearly tri-diagonal and show the state 2 distinctively.  The necessary Markovianity criterion given in Eq. (2) of \cite{NJP2018} is fulfilled. The equilibrium distributions corresponding to (a) and (b) are (0.237, 0.073, 0.285, 0.277, 0.129) and (0.212, 0.069, 0.193, 0.128, 0.270, 0.127) respectively.}
\label{fig:3}
\end{figure}

At this point we could go two ways. Either explore further properties of the states and their transitions or try an economic explanation. The latter is at the margin of our knowledge and thus we further explore the unusual dynamics we encounter. Next step is to look at the transition matrices, as shown in Figure \ref{fig:3}.  {  The transition matrices are nearly tri-diagonal and show the COVID state distinctively.  The necessary Markovianity criterion given in Eq. (2) of \cite{NJP2018} is fulfilled. The equilibrium distributions corresponding to Figs. \ref{fig:3} (a) and (b) are (0.237, 0.073, 0.285, 0.277, 0.129) and (0.212, 0.069, 0.193, 0.128, 0.270, 0.127) respectively.} Note that state 2 has few transitions as can be seen from Fig. \ref{fig:1}, a signature we have never found before.  This reinforces the interest in the {  COVID} state.  The transitions are principally located at the edges of state 2, which indicates that it is essentially a smooth transition.   It is important to mention that for risk assessment,  noise suppression { techniques applied to the correlation matrix rather than to time series} \cite{Guhr_2003, Guhr_2010, THS_2013} are important. 

\subsection*{Distribution of correlation matrix elements over total time horizon}

We note that this anomaly appears during the main COVID period and we may suspect that it is hidden by the panic at the beginning of this pandemia which implies high correlations as we can again see from Fig.  \ref{fig:1}. Therefore, we will look at the relative correlations $RC$ with respect to the S\&P 500 index as defined in Eq. \eqref{eq:2} to explore if these display features of state 2 also during the panic period at the beginning or the slump of S\&P 500 at the beginning of the pandemic. We inspected results \cite{RC2023} obtained in this context previously and distribution of the correlation matrix elements for each epoch turns out to be of particular interest. 

\begin{figure}
           \includegraphics[width=13.5cm]{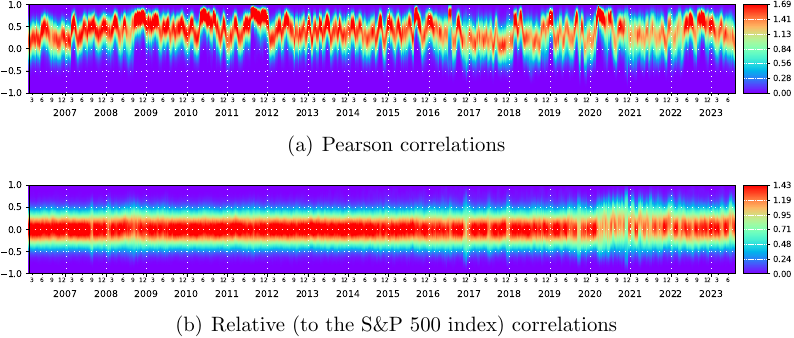}
 \caption{Time evolution of distribution of correlation matrix elements corresponding to (a) Pearson correlation coefficients defined by Eq. \eqref{eq:2} and (b) relative (with respect to the S\&P 500 index) correlation coefficients defined by Eq. \eqref{eq:3}.  For the time period starting June 1st 2020 where state number 2 starts, the fluctuation of the matrix elements of $C$ becomes much faster and does not stop at the end of state 2 (February 1st 2022) but persists. Also, this behavior does not start at the beginning of COVID period but at the beginning of state 2.  Whereas with $RC$, the state starts in March 2020.}
\label{fig:4}
\end{figure}

We found striking results when looking at the histograms for distribution of correlation matrix elements for each of the epochs as shown in Figure \ref{fig:4}.  A more detailed analysis will be given in \cite{RC2023} but a simple ocular inspection shows two points: For the time period starting June 1st 2020 where state number 2 starts, the  fluctuation of the matrix elements become much faster and this first sight behavior does not stop at the end of state 2 but persists. This behavior does not start at the beginning of COVID period but at the beginning of state 2. Indeed it starts when the crash of the stock market and strong following fluctuations approximately end due to panic that can be seen between November 9th 2020 and February 1st 2022 can override the COVID influence to some extent. We therefore relate this to data we obtained in an almost concluded analysis \cite{RC2023} of relative \cite{KenStu, ANZJS2004} and reduced \cite{Heckens_2020, JSM2022, PhyA2022} correlation matrices. We see markedly different behavior that with beginning of COVID, we have a change in the behavior of market as long as the highest correlation does not dictate it.

{  
\subsection*{Participation ratios}

\begin{figure}
           \includegraphics[width=13.5cm]{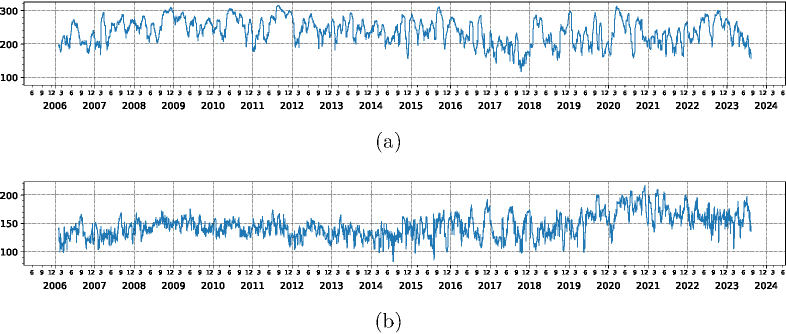}
 \caption{Time evolution of PR defined in Eq. \eqref{eq-pr} for eigenvector of the largest eigenvalue corresponding to (a) Pearson correlation coefficients defined by Eq. \eqref{eq:2} and (b) relative (with respect to the S\&P 500 index) correlation coefficients defined by Eq. \eqref{eq:3}.  The vertical shadowed stripes indicate market crash periods which are usually mentioned in the literature.}
\label{fig:5}
\end{figure}

Participation ratios (PR) gives the number of components that participate significantly in each eigenvector $v$,
\begin{equation}
PR_v = \left[ \displaystyle\sum_{i = 1}^N \left| v_i \right|^4\right] \;.
\label{eq-pr}
\end{equation}
PR takes values between 1 and $N$ and for a Gaussian Orthogonal Ensemble (GOE) has the limiting value of $N/3$ \cite{Brody-RMT, Phys-Rep-Guhr, JPC-2023}.  This GOE result holds true for correlation matrices as well and will be seen in the center of the spectrum for sufficiently long epochs.  We analyze the time evolution of PR corresponding to the largest eigenvalues using Pearson and relative (with respect to the S\&P 500 index) correlation coefficients respectively in Figs. \ref{fig:5} (a) and (b).  For the Pearson correlations, the PR is above the GOE threshold (107.33) for all the epochs. However,  for the relative correlations,  in the COVID period, the PR is constantly high in comparison to other epochs.

\begin{figure}
          \includegraphics[width=13.5cm]{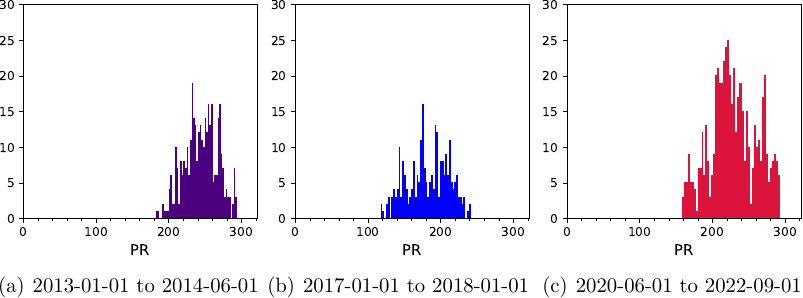}
 \caption{Distribution of PR defined in Eq. \eqref{eq-pr} for eigenvector of the largest eigenvalue corresponding to Pearson correlation coefficients defined by Eq. \eqref{eq:2} for three different time periods -  (a) non-calm period,  (b) calm period,  and (c) COVID period.  Note that the format for the dates used is YYYY-MM-DD. The first four moments (average, variance, skewness, kurtosis) are as follows - (a) (244.5, 525.04, -0.13,  -0.48), (b) (182.34,  799.19, -0.17, -0.83), and (c) (228.66, 1008.69, -0.004, -0.65). Note that these figures have same scales.}
\label{fig:6}
\end{figure}

To probe into further details, we choose three different time periods within our time horizon: (a) 2013-01-01 to 2014-06-01, (b) 2017-01-01 to 2018-01-01,  and (c) 2020-06-01 to 2022-09-01, corresponding respectively to non-calm period,  calm period and the COVID period. We then analyze the histograms for PR for these three time periods as shown in Fig. \ref{fig:6}, obtained using Pearson correlation matrices.  The average of the distribution is highest for the non-calm period,  lowest for the calm period,  and intermediate for the COVID period.  However, the variance and skewness are largest for the COVID state.  The distribution of PR in the COVID period is quite symmetrical, unlike the other two time periods chosen.  We also looked at Inverse Participation Ratios (IPR) and the signal is less clear.  This is not surprising as the IPR is used for analytical purposes as these are entire functions and PR is a natural choice for data analysis.  The statistical analysis of eigenvalues is not conclusive due to reduced sample sizes in the shorter time periods.
}

\section*{Conclusions and future outlook}
\label{sec3} 

{  Starting from a multivariate correlation analysis of financial markets using a methodology that resulted in the definition of market states  \cite{SciRep2012} and using the specific techniques proposed in \cite{NJP2018,  arXiv2020}, we found that a previously non-existent market state appears in a time frame closely related to the COVID pandemic.  Expanding the methodology by using also the relative correlations of stocks with respect to the market index, we get results that we hope will give a deeper insight into the concept of market states. The emergence of the ``state of the market'' represented by the largest eigenvalues for both Pearson and relative correlations seems noteworthy}. The stability of the results is confirmed by the corresponding discussions in the supplemental material \cite{SM}. {  For four states, the COVID state is not visible. Going beyond four, we have shown results for 5-8 states that the qualitative behavior of COVID state remains unchanged.  Although we would like to remark that it might split up increasing the states further. We can not use an arbitrary large number of states but we checked up to 12 states that the COVID state is qualitatively unchanged.  Relative correlations show similar behavior.}The temporal coincidence makes us believe that it has to do with economical consequences of the restrictions and the mindset of the population during the COVID pandemic. This idea is fortified by the observation that the full correlation matrix analysis indicates an onset of the {  COVID} state roughly three months after the onset of this state,  at which time panic sales and the corresponding crash associated with high average correlation are over. This state ends in February 2022 with a few points reappearing at the end of our time horizon.  Being at the end of our time horizon these points are not very reliable,  but at any rate we cannot yet distinguish if we are talking of a very specific and time bound reaction to COVID or whether we see a new general market situation. Time might tell.

It is remarkable that the COVID state is clearly marked as the state with the highest average correlation relative to the  S\&P 500 index.  Note though that the beginning of this state for the relative correlations is roughly coinciding with the strong crash of the stock values while this state for the Pearson correlations appears towards the end of COVID crash. This indicates that the high correlation still dominates the market but once the average correlation decreases, other components become important.  These other components are visible if the general market behavior is removed using relative to the S\&P 500 index correlations.  This increases the relevance of the very concept of relative correlation in financial markets and indeed relates also to recently developed concepts of reduced correlations by Guhr and co-workers \cite{Heckens_2020, JSM2022, PhyA2022}. We present more details about these methods and results in Ref.  \cite{RC2023}.   Indeed we hope that this example will help us in our search for relevant parameters in the stock market beyond the highest eigenvalue of the Pearson correlation matrix (essentially equivalent to the average correlation) yet significantly smaller in number than the huge number of matrix elements of the correlation matrix \cite{SecAna2023, CoGr2023}.  We do not use power map \cite{Guhr_2003, Guhr_2010, THS_2013} for noise suppression as we want to emphasize subtler correlations.  This may even lead to the use of an ``anti Power map'' i.e.  with powers smaller than one. 

\section*{Data Availability}

{  Data is available in figshare repository  \url{https://doi.org/10.6084/m9.figshare.25219880.v1}}; downloaded from \url{https://finance.yahoo.com/}.

\section*{Acknowledgements}

The authors are grateful to Fran\c{c}ois Leyvraz, Anirban Chakraborti,  Thomas Gorin, Roberto Mota, Alejandro Ra{\'u}l Hern{\'a}ndez-Montoya and Andr{\'e}s Ra{\'u}l Cruz-Hern{\'a}ndez for their inputs and suggestions.  

\section*{Funding Statement}

{  We acknowledge financial support from Universidad Nacional Autónoma de México (UNAM) Dirección General de Asuntos del Personal Académico (DGAPA) Programa de Apoyo a Proyectos de Investigación e Innovación Tecnológica (PAPIIT) Project IN113620,  IG101122 and IN102620 and Consejo Nacional de Humanidades Ciencias y Technologías (CONAHCYT) Project Fronteras 425854 and Project 254515.  M. V.  acknowledges financial support from CONAHCYT Project Fronteras 10872.  P. M. gratefully acknowledges fellowships from CONAHCYT Project Fronteras and UNAM–DGAPA PAPIIT.  The funders had no role in study design, data collection and analysis, decision to publish, or preparation of the manuscript.}

\section*{Declarations of interest}

{The authors have declared that no competing interests exist.}

\section*{Use of AI tools declaration}

The authors declare they have not used Artificial Intelligence (AI) tools in the creation of this article.


\newpage

\vspace{0.3cm}

\setcounter{page}{1}

\vspace*{0.2in}

\begin{flushleft}

{\Large
\textbf\newline{Supplemental Material:\\ COVID anomaly in the correlation analysis of S\&P 500 market states}
}
\newline
\\
M. Mija{\'i}l Mart{\'i}nez-Ramos\textsuperscript{1},
Manan Vyas\textsuperscript{1*},
Parisa Majari\textsuperscript{1},
Thomas H.  Seligman\textsuperscript{1, 2}
\\
\bigskip
\textbf{1} Instituto de Ciencias Físicas - Universidad Nacional Autónoma de México,  Cuernavaca, 62210,  Morelos,  México
\\
\textbf{2} Centro Internacional de Ciencias AC - UNAM, Avenida Universidad 1001,  UAEM,  Cuernavaca, 62210,  Morelos,  México
\\
\bigskip

* manan@icf.unam.mx (MV)

\end{flushleft}

\vspace{0.6cm}

\setcounter{figure}{0}
\setcounter{equation}{0}
\setcounter{table}{0}

\renewcommand{\thetable}{S\arabic{table}}
\renewcommand{\thefigure}{S\arabic{figure}}
\renewcommand{\theequation}{S\arabic{equation}}
\renewcommand{\thepage}{S\arabic{page}}
\renewcommand{\thesection}{S\arabic{section}}

\begin{figure}
           \includegraphics[width=13.5cm]{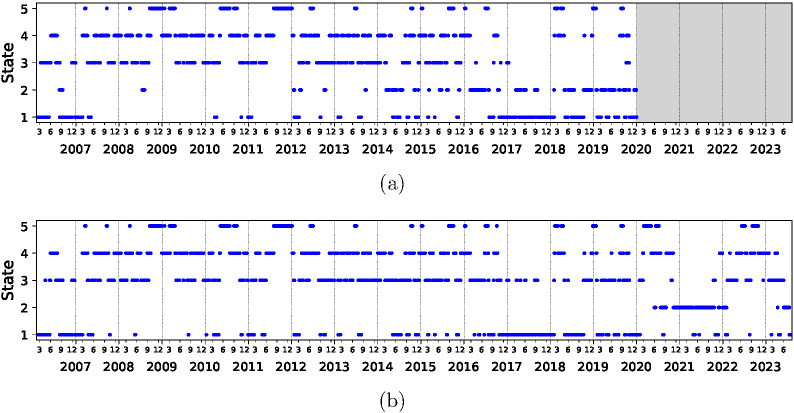}
           \caption{Time evolution of market states of the S\&P 500 data using Pearson correlation coefficients in a time horizon from January 3rd 2006 to (a) December 31st 2019 and (b) August 10th 2023,  with an epoch of 20 trading days.  Pearson correlation matrix elements are computed using logarithmic return time series of adjusted closing prices.  The market states are arranged in order of increasing average correlations.  The average correlations for the market states are (a) 0.16, 0.28, 0.30, 0.43, 0.60 and (b) 0.17, 0.26, 0.30, 0.44, 0.61, respectively.  State 2 corresponding to COVID anomaly shows up in the time period starting June 1, 2020 until February 1, 2022.}
\label{A1}
\end{figure}

\begin{figure}
           \includegraphics[width=13.5cm]{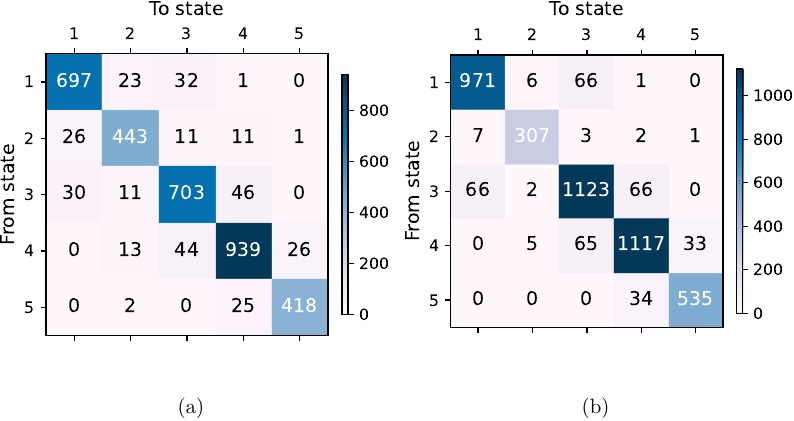}
\caption{Transition matrices corresponding to Pearson correlation coefficients showing transitions between different market states of the S\&P 500 data from January 3rd 2006 to (a) December 31st 2019 and (b) August 10th 2023.  The difference due to appearance of COVID anomaly is visible in transition matrix as well.  Also,  the necessary Markovianity criterion given in Eq. (2) of \cite{NJP2018} is fulfilled.  The equilibrium distributions corresponding to time periods of (a) and (b) are $(0.228, 0.224, 0.131, 0.286, 0.1308)$ and $(0.240, 0.073, 0.285, 0.277, 0.129)$ respectively.}
\label{A2}
\end{figure}

Classification of the S\&P 500 market states into (a) five and (b) six clusters displays the atypical state 2 corresponding to COVID anomaly.  To further substantiate the existence of this state, we compare the time evolution of market states starting January 3rd 2006 to (a) December 31st 2019 and (b) August 10th 2023 in Figures \ref{A1}(a) and \ref{A1}(b) respectively.  Also, the structure of the transition matrices display this signature clearly as shown in Figs. \ref{A2}(a) and \ref{A2}(b). The necessary Markovianity criterion given in Eq. (2) of \cite{S-NJP2018} is also fulfilled by the transition matrices.  The equilibrium distributions are $(0.240, 0.073, 0.285, 0.277, 0.129)$ and $(0.212, 0.069, 0.193, 0.128, 0.270, 0.127)$ respectively for five and six market states.

\begin{figure}
\centering
         \includegraphics[width=6.75cm]{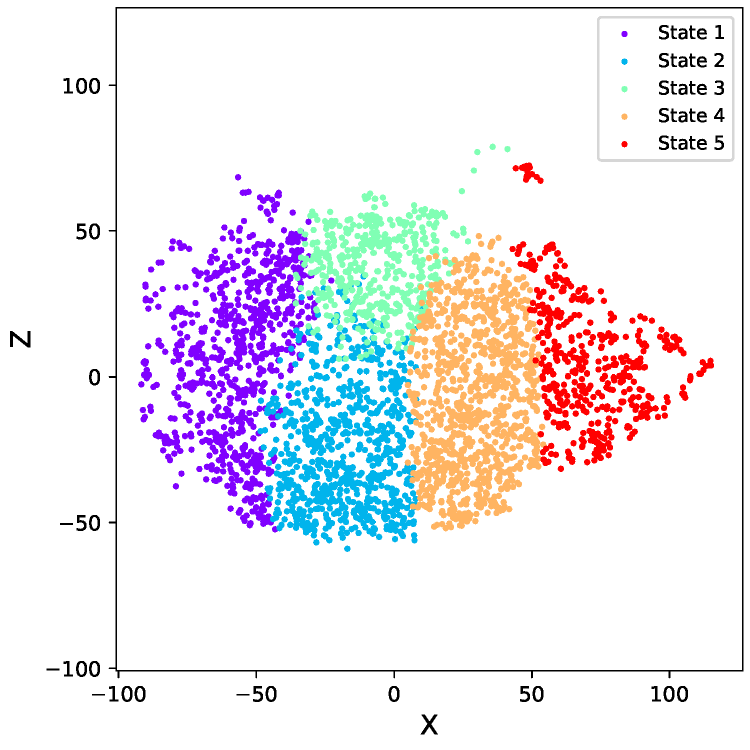}
\caption{$L^1$ 3D Multi Dimensional Scaling after $k$-means clustering of 3503 Pearson correlation matrices (projections on principal components) of the five S\&P 500 market states from January 3rd 2006 to December 31st 2019.  The image above shows a single frame from the video. Click \href{run:Video1.mp4}{here} to play the video. }
\label{video1} 
 \end{figure}

\begin{figure}
\centering
\includegraphics[width=6.75cm]{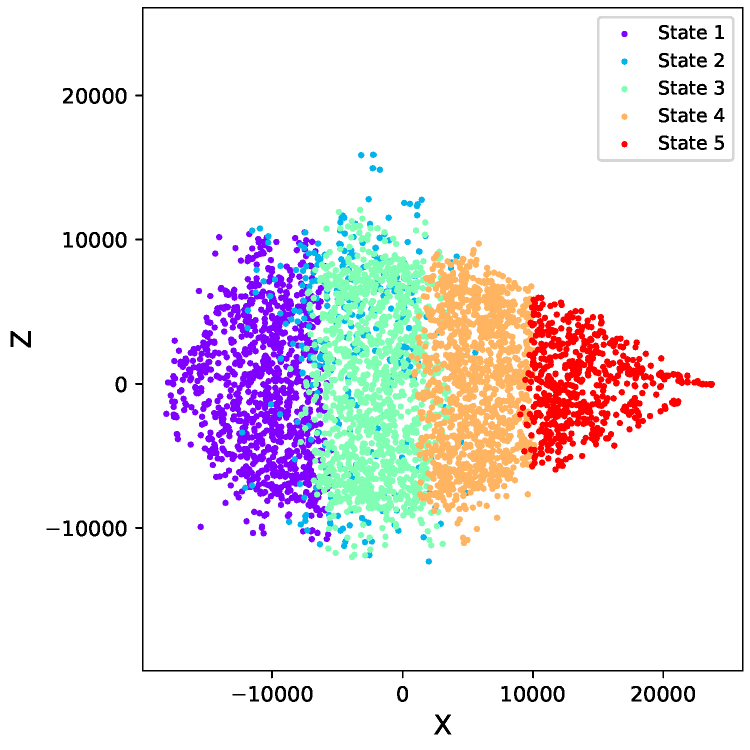}
\caption{$L^1$ 3D Multi Dimensional Scaling after $k$-means clustering of 4411 Pearson correlation matrices (projections on principal components) of the five S\&P 500 market states from January 3rd 2006 to August 10th 2023.  The image above shows a single frame from the video. Click \href{run:Video1.mp4}{here} to play the video.}
\label{video2}
 \end{figure}

\begin{figure}
\centering
\includegraphics[width=6.75cm]{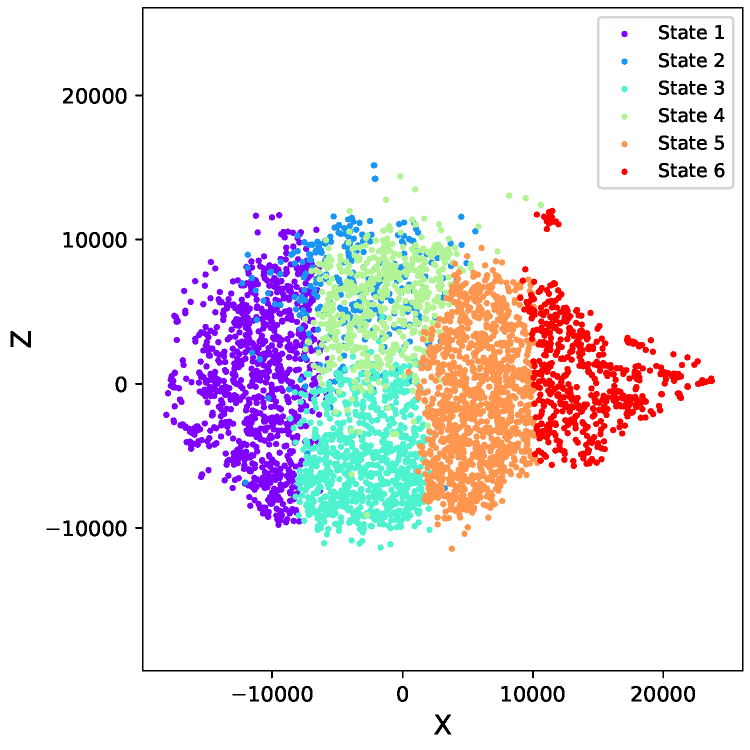}
\caption{$L^1$ 3D Multi Dimensional Scaling after $k$-means clustering of 4411 Pearson correlation states (projections on principal components) of the six S\&P 500 market states from January 3rd 2006 to August 10th 2023.  The image above shows a single frame from the video. Click \href{run:Video3.mp4}{here} to play the video.}
\label{video3}
 \end{figure}

This immediately makes us suspect that the linear alignment seen in the dimensionally scaled picture of all the correlation matrices is no longer conserved \cite{S-NJP2018, S-Chap23}. We therefore show in Figures \ref{video1}, \ref{video2}, and \ref{video3} the Pearson correlation matrices with 5 market states of the S\&P 500 data from January 3rd 2006 to December 31st 2019,  5 market states of the S\&P 500 data from January 3rd 2006 to August 10th 2023, 6 market states of the S\&P 500 data from January 3rd 2006 to August 10th 2023, respectively,  after subjecting them to dimensional scaling according to the recipe given in \cite{S-borg2005modern}.  These figures show a single frame from the video, click in the captions to play the videos.  We scale down to 3 dimensions and we see that the usually rather smooth picture shows a bulge in Figure \ref{video2} which under close scrutiny results to correspond exactly to state 2 as demonstrated by the color code of the picture.  

\begin{figure}
\includegraphics[width=13.5cm]{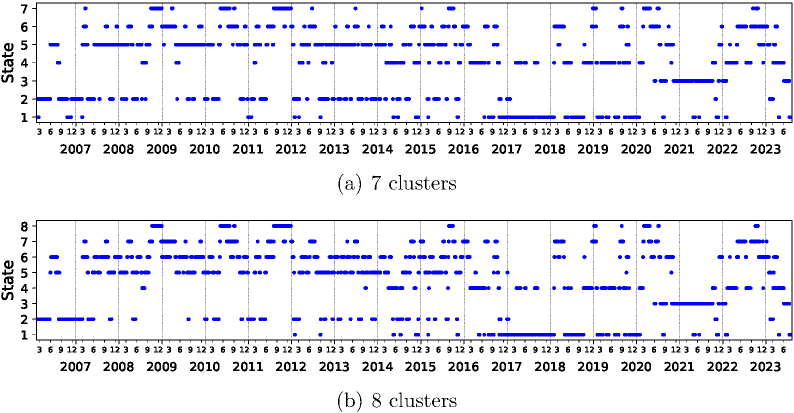}
 \caption{Time evolution of market states  of the S\&P 500 data using Pearson correlation coefficients in a time horizon from January 3rd 2006 to August 10th 2023 for (a) seven and (b) eight clusters.  Pearson correlation matrix elements are computed using logarithmic return time series of adjusted closing prices.  The market states are arranged in order of increasing average correlations.  The average correlations for the market states are (a) 0.15, 0.24, 0.26, 0.30,  0.39, 0.49, 0.65 and (b) 0.15,  0.20,  0.26,  0.30,  0.31,  0.42,  0.52,  0.66, respectively. }
\label{A4}
\end{figure}

We also analyze the existence of state corresponding to COVID anomaly by increasing the number of clusters and the results are as shown in Fig. \ref{A4}.  Although we show results only for seven and eight clusters, we have verified that this state exists for $9, 10, 11$ and 12 clusters as well. This illustrates that the atypical state corresponding to COVID is stable.

\begin{thebibliography}{99}

\bibitem{SciRep2012}
M{\"u}nnix MC, Shimada T, Sch{\"a}fer R, Leyvraz F, Seligman TH, Guhr T, et~al.
\newblock Identifying States of a Financial Market.
\newblock Scientific Reports. 2012;2(1):644.
\newblock Available from: \url{https://doi.org/10.1038/srep00644}.

\bibitem{Zhou_2018}
Zhou L, Qiu L, Gu C, Yang H.
\newblock Immediate causality network of stock markets.
\newblock Europhysics Letters. 2018 apr;121(4):48002.
\newblock Available from: \url{https://dx.doi.org/10.1209/0295-5075/121/48002}.

\bibitem{PhysRevE.97.052312}
Anand K, Khedair J, K\"uhn R.
\newblock Structural model for fluctuations in financial markets.
\newblock Phys Rev E. 2018 May;97:052312.
\newblock Available from:
  \url{https://link.aps.org/doi/10.1103/PhysRevE.97.052312}.

\bibitem{Tang_2018}
Tang J, Khoja L, Heinimann HR.
\newblock Characterisation of survivability resilience with dynamic stock
  interdependence in financial networks.
\newblock Applied Network Science. 2018;3:23.
\newblock Available from: \url{https://doi.org/10.1007/s41109-018-0086-z}.

\bibitem{NJP2018}
Pharasi HK, Sharma K, Chatterjee R, Chakraborti A, Leyvraz F, Seligman TH.
\newblock Identifying Long-Term Precursors of Financial Market Crashes Using
  Correlation Patterns.
\newblock New J Phys. 2018;20(10):103041.
{  \newblock Available from: \url{https://dx.doi.org/10.1088/1367-2630/aae7e0}.}

\bibitem{Springer2019}
Pharasi HK, Sharma K, Chakraborti A, Seligman TH.
\newblock In: Abergel F, Chakrabarti BK, Chakraborti A, Deo N, Sharma K,
  editors. Complex Market Dynamics in the Light of Random Matrix Theory. Cham:
  Springer International Publishing; 2019. p. 13-34.
\newblock Available from: \url{https://doi.org/10.1007/978-3-030-11364-3_2}.

{ 
\bibitem{arXiv2020}
Pharasi HK, Seligman E, Sadhukhan S, Majari P,  Seligman TH.
\newblock Dynamics of market states and risk assessment.
\newblock Physica A. 2024;633:129396.
\newblock Available from: \url{https://doi.org/10.1016/j.physa.2023.129396}.
}
\bibitem{Wang_2020}
Wang S, Gartzke S, Schreckenberg M, Guhr T.
\newblock Quasi-stationary states in temporal correlations for traffic systems:
  Cologne orbital motorway as an example.
\newblock Journal of Statistical Mechanics: Theory and Experiment. 2020
  oct;2020(10):103404.
\newblock Available from: \url{https://dx.doi.org/10.1088/1742-5468/abbcd3}.

\bibitem{Nie_2020}
Nie CX.
\newblock A network-based method for detecting critical events of correlation
  dynamics in financial markets.
\newblock Europhysics Letters. 2020 sep;131(5):50001.
\newblock Available from: \url{https://dx.doi.org/10.1209/0295-5075/131/50001}.

\bibitem{Heckens_2020}
Heckens AJ, Krause SM, Guhr T.
\newblock Uncovering the dynamics of correlation structures relative to the
  collective market motion.
\newblock Journal of Statistical Mechanics: Theory and Experiment. 2020
  oct;2020(10):103402.
\newblock Available from: \url{https://dx.doi.org/10.1088/1742-5468/abb6e2}.

\bibitem{JAMES-2022}
James N, Menzies M, Chin K.
\newblock Economic state classification and portfolio optimisation with
  application to stagflationary environments.
\newblock Chaos, Solitons \& Fractals. 2022;164:112664.
\newblock Available from:
  \url{https://www.sciencedirect.com/science/article/pii/S0960077922008438}.

\bibitem{JSM2022}
Heckens AJ, Guhr T.
\newblock A new attempt to identify long-term precursors for endogenous
  financial crises in the market correlation structures.
\newblock Journal of Statistical Mechanics: Theory and Experiment. 2022
  apr;2022(4):043401.
\newblock Available from: \url{https://dx.doi.org/10.1088/1742-5468/ac59ab}.

\bibitem{PhyA2022}
Heckens AJ, Guhr T.
\newblock New collectivity measures for financial covariances and correlations.
\newblock Physica A: Statistical Mechanics and its Applications.
  2022;604:127704.
\newblock Available from:
  \url{https://www.sciencedirect.com/science/article/pii/S0378437122004666}.

\bibitem{MSBook}
Mantegna RN, Stanley HE.
\newblock An Introduction to Econophysics: Correlations and Complexity in
  Finance.
\newblock {Cambridge University Press}; 2000.

\bibitem{PRL1999}
Laloux L, Cizeau P, Bouchaud JP, Potters M.
\newblock Noise {{Dressing}} of {{Financial Correlation Matrices}}.
\newblock Phys Rev Lett. 1999;83(7):1467-70.
{  \newblock Available from: \url{https://doi.org/10.1103/PhysRevLett.83.1467}.}

\bibitem{Chap23}
Pharasi HK, Sadhukhan S, Majari P, Chakraborti A, Seligman TH.
\newblock Market state dynamics in correlation matrix space.
\newblock In: Quantum decision theory and complexity modelling in economics and
  public policy. Springer; in press 2023. 

\bibitem{RC2023}
Mart{\'i}nez-Ramos MM, Majari P, Vyas M.
\newblock Correlations beyond the trend: A financial market analysis. in prep
  (2023).

\bibitem{KenStu}
Kendall MG, Stuart A.
\newblock The Advanced Theory of Statistics, Volume 2: Inference and
  Relationship.
\newblock {Griffin}; 1973.

\bibitem{ANZJS2004}
Baba K, Shibata R, Sibuya M.
\newblock Partial correlation and conditional correlation as measures of
  conditional independence.
\newblock Australian \& New Zealand Journal of Statistics. 2004;46(4):657-64.
\newblock Available from:
  \url{https://onlinelibrary.wiley.com/doi/abs/10.1111/j.1467-842X.2004.00360.x}.

{ 
\bibitem{SecAna2023}
Mart{\'i}nez-Ramos MM,  Majari P, Cruz-Hernández A. R., Pharasi H. K. , Vyas M.
\newblock Coarse graining correlation matrices according to macrostructures: Financial markets as a paradigm. 
\newblock Available from:
  \url{https://arxiv.org/pdf/2402.05364.pdf}.
}
\bibitem{CoGr2023}
Mart{\'i}nez-Ramos MM, Vyas M, Majari P, Seligman TH.
\newblock Coarse graining correlation matrices with applications in financial
  markets. in prep (2023).

\bibitem{Guhr_2003}
Guhr T, Kälber B.
\newblock A new method to estimate the noise in financial correlation matrices.
\newblock Journal of Physics A: Mathematical and General. 2003 mar;36(12):3009.
\newblock Available from: \url{https://dx.doi.org/10.1088/0305-4470/36/12/310}.

\bibitem{Guhr_2010}
Schäfer R, Guhr T.
\newblock {Local normalization: Uncovering correlations in non-stationary
  financial time series}.
\newblock Physica A: Statistical Mechanics and its Applications.
  2010;389(18):3856-65.
\newblock Available from:
  \url{https://ideas.repec.org/a/eee/phsmap/v389y2010i18p3856-3865.html}.

\bibitem{THS_2013}
Vinayak, Sch\"afer R, Seligman TH.
\newblock Emerging spectra of singular correlation matrices under small
  power-map deformations.
\newblock Phys Rev E. 2013 Sep;88:032115.
\newblock Available from:
  \url{https://link.aps.org/doi/10.1103/PhysRevE.88.032115}.

\bibitem{WHO}
WHO Director-General's opening remarks at the media briefing on COVID-19—11
  March 2020; March 2020.
\newblock
  \url{https://www.who.int/dg/speeches/detail/who-director-general-s-opening-remarks-at-the-media-briefing-on-covid-19---11-march-2020/}.

\bibitem{SM}
Mart{\'i}nez-Ramos MM, Vyas M, Majari P, Seligman TH. Supplemental Material:
  COVID anomaly in the correlation analysis of S\&P 500 market states; 2023.
\newblock \url{URL_will_be_inserted_by_publisher}.

\bibitem{yf}
2023 Yahoo finance database; accessed on 11 August, 2023 for S\&P 500.
\newblock \url{https://finance.yahoo.com/}.

\bibitem{borg2005modern}
Borg I, Groenen PJF.
\newblock Modern Multidimensional Scaling: Theory and Applications.
\newblock Springer Series in Statistics. Springer New York; 2005.
\newblock Available from:
  \url{https://books.google.com.mx/books?id=duTODldZzRcC}.

{ 
\bibitem{Brody-RMT}
Brody T. A., Flores J., French J. B., Mello P.A., Pandey A., Wong S. S. M.
\newblock Random-matrix physics: Spectrum and strength fluctuations.
\newblock Reviews of Modern Physics. 1981;53(3):385-479.
\newblock Available from:
  \url{https://doi.org/10.1103/RevModPhys.53.385}.

\bibitem{Phys-Rep-Guhr}
Guhr T., Mueller-Groeling A., Weidenmueller H. A.
\newblock Random-matrix theories in quantum physics: common concepts.
\newblock Physics Reports. 1998;299:189-425.
\newblock Available from:
  \url{https://doi.org/10.1016/S0370-1573(97)00088-4}.

\bibitem{JPC-2023}
Salgado-Hernández J. E., Vyas M.
\newblock Non-linear correlation analysis in financial markets using hierarchical clustering.
\newblock Journal of Physics Communications. 2023;7:055003.
\newblock Available from:
  \url{https://dx.doi.org/10.1088/2399-6528/acd618}.
}
\end{thebibliography}

\newpage
\begin{thebibliography}{1}
\expandafter\ifx\csname url\endcsname\relax
  \def\url#1{\texttt{#1}}\fi
\expandafter\ifx\csname urlprefix\endcsname\relax\def\urlprefix{URL }\fi
\expandafter\ifx\csname href\endcsname\relax
  \def\href#1#2{#2} \def\path#1{#1}\fi

\bibitem{S-NJP2018}
H.~K. Pharasi, K.~Sharma, R.~Chatterjee, A.~Chakraborti, F.~Leyvraz, T.~H.
  Seligman, Identifying long-term precursors of financial market crashes using
  correlation patterns, New J. Phys. 20~(10) (2018) 103041.
\newblock \href {https://doi.org/10.1088/1367-2630/aae7e0}
  {\path{doi:10.1088/1367-2630/aae7e0}}.

\bibitem{S-Chap23}
H.~K. Pharasi, S.~Sadhukhan, P.~Majari, A.~Chakraborti, T.~H. Seligman, Market
  state dynamics in correlation matrix space, in: Quantum decision theory and
  complexity modelling in economics and public policy, Springer, in press 2023.

\bibitem{S-borg2005modern}
I.~Borg, P.~Groenen,
  \href{https://books.google.com.mx/books?id=duTODldZzRcC}{Modern
  Multidimensional Scaling: Theory and Applications}, Springer Series in
  Statistics, Springer New York, 2005.
\newline\urlprefix\url{https://books.google.com.mx/books?id=duTODldZzRcC}

\end{thebibliography}

\newpage

{\bf List of the 322 stocks analyzed}

{\tiny
\begin{longtable}[!p]{lll}
\toprule
\textbf{Sector} & \textbf{Ticker} & \textbf{Name}  \\
\midrule
Basic Materials & APD & Air Products and Chemicals, Inc. \\
Basic Materials & CF & CF Industries Holdings, Inc. \\
Basic Materials & ECL & Ecolab Inc. \\
Basic Materials & FCX & Freeport-McMoRan Inc. \\
Basic Materials & FMC & FMC Corporation \\
Basic Materials & IFF & International Flavors \& Fragrances Inc. \\
Basic Materials & MOS & The Mosaic Company \\
Basic Materials & NEM & Newmont Corporation \\
Basic Materials & NUE & Nucor Corporation \\
Basic Materials & PPG & PPG Industries, Inc. \\
Basic Materials & SHW & The Sherwin-Williams Company \\
Basic Materials & VMC & Vulcan Materials Company \\
Communication Services & ATVI & Activision Blizzard, Inc. \\
Communication Services & CMCSA & Comcast Corporation \\
Communication Services & DISH & DISH Network Corporation \\
Communication Services & EA & Electronic Arts Inc. \\
Communication Services & GOOG & Alphabet Inc. \\
Communication Services & GOOGL & Alphabet Inc. \\
Communication Services & IPG & The Interpublic Group of Companies, Inc. \\
Communication Services & NFLX & Netflix, Inc. \\
Communication Services & OMC & Omnicom Group Inc. \\
Communication Services & T & AT\&T Inc. \\
Communication Services & TTWO & Take-Two Interactive Software, Inc. \\
Communication Services & VZ & Verizon Communications Inc. \\
Consumer Cyclical & AAP & Advance Auto Parts, Inc. \\
Consumer Cyclical & AMZN & Amazon.com, Inc. \\
Consumer Cyclical & AVY & Avery Dennison Corporation \\
Consumer Cyclical & AZO & AutoZone, Inc. \\
Consumer Cyclical & BBY & Best Buy Co., Inc. \\
Consumer Cyclical & BKNG & Booking Holdings Inc. \\
Consumer Cyclical & CCL & Carnival Corporation \& plc \\
Consumer Cyclical & DHI & D.R. Horton, Inc. \\
Consumer Cyclical & EBAY & eBay Inc. \\
Consumer Cyclical & EXPE & Expedia Group, Inc. \\
Consumer Cyclical & F & Ford Motor Company \\
Consumer Cyclical & GPC & Genuine Parts Company \\
Consumer Cyclical & GPS & The Gap, Inc. \\
Consumer Cyclical & HAS & Hasbro, Inc. \\
Consumer Cyclical & HD & The Home Depot, Inc. \\
Consumer Cyclical & HOG & Harley-Davidson, Inc. \\
Consumer Cyclical & HRB & H\&R Block, Inc. \\
Consumer Cyclical & IP & International Paper Company \\
Consumer Cyclical & JWN & Nordstrom, Inc. \\
Consumer Cyclical & KMX & CarMax, Inc. \\
Consumer Cyclical & KSS & Kohl's Corporation \\
Consumer Cyclical & LEG & Leggett \& Platt, Incorporated \\
Consumer Cyclical & LEN & Lennar Corporation \\
Consumer Cyclical & LKQ & LKQ Corporation \\
Consumer Cyclical & LOW & Lowe's Companies, Inc. \\
Consumer Cyclical & M & Macy's, Inc. \\
Consumer Cyclical & MAR & Marriott International, Inc. \\
Consumer Cyclical & MCD & McDonald's Corporation \\
Consumer Cyclical & MGM & MGM Resorts International \\
Consumer Cyclical & MHK & Mohawk Industries, Inc. \\
Consumer Cyclical & NKE & NIKE, Inc. \\
Consumer Cyclical & ORLY & O'Reilly Automotive, Inc. \\
Consumer Cyclical & PHM & PulteGroup, Inc. \\
Consumer Cyclical & PKG & Packaging Corporation of America \\
Consumer Cyclical & PVH & PVH Corp. \\
Consumer Cyclical & RL & Ralph Lauren Corporation \\
Consumer Cyclical & ROST & Ross Stores, Inc. \\
Consumer Cyclical & SBUX & Starbucks Corporation \\
Consumer Cyclical & SEE & Sealed Air Corporation \\
Consumer Cyclical & TJX & The TJX Companies, Inc. \\
Consumer Cyclical & TPR & Tapestry, Inc. \\
Consumer Cyclical & UAA & Under Armour, Inc. \\
Consumer Cyclical & VFC & V.F. Corporation \\
Consumer Cyclical & WHR & Whirlpool Corporation \\
Consumer Cyclical & WYNN & Wynn Resorts, Limited \\
Consumer Cyclical & YUM & Yum! Brands, Inc. \\
Consumer Defensive & ADM & Archer-Daniels-Midland Company \\
Consumer Defensive & CAG & Conagra Brands, Inc. \\
Consumer Defensive & CHD & Church \& Dwight Co., Inc. \\
Consumer Defensive & CL & Colgate-Palmolive Company \\
Consumer Defensive & CLX & The Clorox Company \\
Consumer Defensive & COST & Costco Wholesale Corporation \\
Consumer Defensive & CPB & Campbell Soup Company \\
Consumer Defensive & DLTR & Dollar Tree, Inc. \\
Consumer Defensive & EL & The Estée Lauder Companies Inc. \\
Consumer Defensive & GIS & General Mills, Inc. \\
Consumer Defensive & HRL & Hormel Foods Corporation \\
Consumer Defensive & HSY & The Hershey Company \\
Consumer Defensive & K & Kellogg Company \\
Consumer Defensive & KMB & Kimberly-Clark Corporation \\
Consumer Defensive & KO & The Coca-Cola Company \\
Consumer Defensive & KR & The Kroger Co. \\
Consumer Defensive & MDLZ & Mondelez International, Inc. \\
Consumer Defensive & MKC & McCormick \& Company, Incorporated \\
Consumer Defensive & MNST & Monster Beverage Corporation \\
Consumer Defensive & MO & Altria Group, Inc. \\
Consumer Defensive & NWL & Newell Brands Inc. \\
Consumer Defensive & PEP & PepsiCo, Inc. \\
Consumer Defensive & PG & The Procter \& Gamble Company \\
Consumer Defensive & SJM & The J. M. Smucker Company \\
Consumer Defensive & STZ & Constellation Brands, Inc. \\
Consumer Defensive & SYY & Sysco Corporation \\
Consumer Defensive & TAP & Molson Coors Beverage Company \\
Consumer Defensive & TGT & Target Corporation \\
Consumer Defensive & TSN & Tyson Foods, Inc. \\
Consumer Defensive & WMT & Walmart Inc. \\
Energy & APA & APA Corporation \\
Energy & COP & ConocoPhillips \\
Energy & CVX & Chevron Corporation \\
Energy & DVN & Devon Energy Corporation \\
Energy & EOG & EOG Resources, Inc. \\
Energy & FTI & TechnipFMC plc \\
Energy & HAL & Halliburton Company \\
Energy & HES & Hess Corporation \\
Energy & HP & Helmerich \& Payne, Inc. \\
Energy & MRO & Marathon Oil Corporation \\
Energy & NOV & NOV Inc. \\
Energy & OKE & ONEOK, Inc. \\
Energy & PXD & Pioneer Natural Resources Company \\
Energy & SLB & Schlumberger Limited \\
Energy & VLO & Valero Energy Corporation \\
Energy & WMB & The Williams Companies, Inc. \\
Energy & XOM & Exxon Mobil Corporation \\
Financial Services & AFL & Aflac Incorporated \\
Financial Services & AIG & American International Group, Inc. \\
Financial Services & AIZ & Assurant, Inc. \\
Financial Services & AJG & Arthur J. Gallagher \& Co. \\
Financial Services & AMG & Affiliated Managers Group, Inc. \\
Financial Services & AMP & Ameriprise Financial, Inc. \\
Financial Services & AON & Aon plc \\
Financial Services & AXP & American Express Company \\
Financial Services & BAC & Bank of America Corporation \\
Financial Services & BEN & Franklin Resources, Inc. \\
Financial Services & BK & The Bank of New York Mellon Corporation \\
Financial Services & BLK & BlackRock, Inc. \\
Financial Services & C & Citigroup Inc. \\
Financial Services & CINF & Cincinnati Financial Corporation \\
Financial Services & CMA & Comerica Incorporated \\
Financial Services & CME & CME Group Inc. \\
Financial Services & FITB & Fifth Third Bancorp \\
Financial Services & GS & The Goldman Sachs Group, Inc. \\
Financial Services & HBAN & Huntington Bancshares Incorporated \\
Financial Services & HIG & The Hartford Financial Services Group, Inc. \\
Financial Services & ICE & Intercontinental Exchange, Inc. \\
Financial Services & IVZ & Invesco Ltd. \\
Financial Services & JPM & JPMorgan Chase \& Co. \\
Financial Services & KEY & KeyCorp \\
Financial Services & L & Loews Corporation \\
Financial Services & LNC & Lincoln National Corporation \\
Financial Services & MCO & Moody's Corporation \\
Financial Services & MET & MetLife, Inc. \\
Financial Services & MMC & Marsh \& McLennan Companies, Inc. \\
Financial Services & MS & Morgan Stanley \\
Financial Services & MTB & M\&T Bank Corporation \\
Financial Services & NDAQ & Nasdaq, Inc. \\
Financial Services & NTRS & Northern Trust Corporation \\
Financial Services & PFG & Principal Financial Group, Inc. \\
Financial Services & PGR & The Progressive Corporation \\
Financial Services & PNC & The PNC Financial Services Group, Inc. \\
Financial Services & PRU & Prudential Financial, Inc. \\
Financial Services & RF & Regions Financial Corporation \\
Financial Services & RJF & Raymond James Financial, Inc. \\
Financial Services & SCHW & The Charles Schwab Corporation \\
Financial Services & SPGI & S\&P Global Inc. \\
Financial Services & STT & State Street Corporation \\
Financial Services & TROW & T. Rowe Price Group, Inc. \\
Financial Services & TRV & The Travelers Companies, Inc. \\
Financial Services & UNM & Unum Group \\
Financial Services & USB & U.S. Bancorp \\
Financial Services & WFC & Wells Fargo \& Company \\
Financial Services & ZION & Zions Bancorporation, National Association \\
Healthcare & A & Agilent Technologies, Inc. \\
Healthcare & ABC & AmerisourceBergen Corporation \\
Healthcare & ABT & Abbott Laboratories \\
Healthcare & ALGN & Align Technology, Inc. \\
Healthcare & AMGN & Amgen Inc. \\
Healthcare & BAX & Baxter International Inc. \\
Healthcare & BDX & Becton, Dickinson and Company \\
Healthcare & BIIB & Biogen Inc. \\
Healthcare & BMY & Bristol-Myers Squibb Company \\
Healthcare & BSX & Boston Scientific Corporation \\
Healthcare & CI & The Cigna Group \\
Healthcare & CNC & Centene Corporation \\
Healthcare & COO & The Cooper Companies, Inc. \\
Healthcare & CVS & CVS Health Corporation \\
Healthcare & DGX & Quest Diagnostics Incorporated \\
Healthcare & DVA & DaVita Inc. \\
Healthcare & EW & Edwards Lifesciences Corporation \\
Healthcare & GILD & Gilead Sciences, Inc. \\
Healthcare & HOLX & Hologic, Inc. \\
Healthcare & HSIC & Henry Schein, Inc. \\
Healthcare & HUM & Humana Inc. \\
Healthcare & IDXX & IDEXX Laboratories, Inc. \\
Healthcare & ILMN & Illumina, Inc. \\
Healthcare & INCY & Incyte Corporation \\
Healthcare & ISRG & Intuitive Surgical, Inc. \\
Healthcare & JNJ & Johnson \& Johnson \\
Healthcare & LH & Laboratory Corporation of America Holdings \\
Healthcare & LLY & Eli Lilly and Company \\
Healthcare & MDT & Medtronic plc \\
Healthcare & MRK & Merck \& Co., Inc. \\
Healthcare & MTD & Mettler-Toledo International Inc. \\
Healthcare & PFE & Pfizer Inc. \\
Healthcare & PRGO & Perrigo Company plc \\
Healthcare & REGN & Regeneron Pharmaceuticals, Inc. \\
Healthcare & RMD & ResMed Inc. \\
Healthcare & SYK & Stryker Corporation \\
Healthcare & TMO & Thermo Fisher Scientific Inc. \\
Healthcare & UHS & Universal Health Services, Inc. \\
Healthcare & UNH & UnitedHealth Group Incorporated \\
Healthcare & VRTX & Vertex Pharmaceuticals Incorporated \\
Healthcare & WAT & Waters Corporation \\
Healthcare & WBA & Walgreens Boots Alliance, Inc. \\
Healthcare & XRAY & DENTSPLY SIRONA Inc. \\
Healthcare & ZBH & Zimmer Biomet Holdings, Inc. \\
Industrials & AAL & American Airlines Group Inc. \\
Industrials & ADP & Automatic Data Processing, Inc. \\
Industrials & ALK & Alaska Air Group, Inc. \\
Industrials & AME & AMETEK, Inc. \\
Industrials & AOS & A. O. Smith Corporation \\
Industrials & BA & The Boeing Company \\
Industrials & CAT & Caterpillar Inc. \\
Industrials & CHRW & C.H. Robinson Worldwide, Inc. \\
Industrials & CMI & Cummins Inc. \\
Industrials & CSX & CSX Corporation \\
Industrials & CTAS & Cintas Corporation \\
Industrials & DE & Deere \& Company \\
Industrials & DOV & Dover Corporation \\
Industrials & EFX & Equifax Inc. \\
Industrials & EMR & Emerson Electric Co. \\
Industrials & ETN & Eaton Corporation plc \\
Industrials & EXPD & Expeditors International of Washington, Inc. \\
Industrials & FAST & Fastenal Company \\
Industrials & FDX & FedEx Corporation \\
Industrials & FLS & Flowserve Corporation \\
Industrials & GD & General Dynamics Corporation \\
Industrials & GE & General Electric Company \\
Industrials & GPN & Global Payments Inc. \\
Industrials & GWW & W.W. Grainger, Inc. \\
Industrials & ITW & Illinois Tool Works Inc. \\
Industrials & JBHT & J.B. Hunt Transport Services, Inc. \\
Industrials & JCI & Johnson Controls International plc \\
Industrials & LMT & Lockheed Martin Corporation \\
Industrials & LUV & Southwest Airlines Co. \\
Industrials & MAS & Masco Corporation \\
Industrials & NOC & Northrop Grumman Corporation \\
Industrials & NSC & Norfolk Southern Corporation \\
Industrials & PAYX & Paychex, Inc. \\
Industrials & PCAR & PACCAR Inc \\
Industrials & PH & Parker-Hannifin Corporation \\
Industrials & PNR & Pentair plc \\
Industrials & PWR & Quanta Services, Inc. \\
Industrials & RHI & Robert Half Inc. \\
Industrials & ROK & Rockwell Automation, Inc. \\
Industrials & RSG & Republic Services, Inc. \\
Industrials & SWK & Stanley Black \& Decker, Inc. \\
Industrials & TXT & Textron Inc. \\
Industrials & UNP & Union Pacific Corporation \\
Industrials & UPS & United Parcel Service, Inc. \\
Industrials & URI & United Rentals, Inc. \\
Industrials & WM & Waste Management, Inc. \\
Real Estate & O & Realty Income Corporation \\
Technology & AAPL & Apple Inc. \\
Technology & ACN & Accenture plc \\
Technology & ADBE & Adobe Inc. \\
Technology & ADI & Analog Devices, Inc. \\
Technology & ADSK & Autodesk, Inc. \\
Technology & AKAM & Akamai Technologies, Inc. \\
Technology & AMAT & Applied Materials, Inc. \\
Technology & AMD & Advanced Micro Devices, Inc. \\
Technology & ANSS & ANSYS, Inc. \\
Technology & APH & Amphenol Corporation \\
Technology & CDNS & Cadence Design Systems, Inc. \\
Technology & CRM & Salesforce, Inc. \\
Technology & CSCO & Cisco Systems, Inc. \\
Technology & CTSH & Cognizant Technology Solutions Corporation \\
Technology & DXC & DXC Technology Company \\
Technology & FFIV & F5, Inc. \\
Technology & FIS & Fidelity National Information Services, Inc. \\
Technology & GLW & Corning Incorporated \\
Technology & GRMN & Garmin Ltd. \\
Technology & HPQ & HP Inc. \\
Technology & IBM & International Business Machines Corporation \\
Technology & INTC & Intel Corporation \\
Technology & INTU & Intuit Inc. \\
Technology & IT & Gartner, Inc. \\
Technology & JNPR & Juniper Networks, Inc. \\
Technology & KLAC & KLA Corporation \\
Technology & LRCX & Lam Research Corporation \\
Technology & MCHP & Microchip Technology Incorporated \\
Technology & MSFT & Microsoft Corporation \\
Technology & MSI & Motorola Solutions, Inc. \\
Technology & MU & Micron Technology, Inc. \\
Technology & NTAP & NetApp, Inc. \\
Technology & NVDA & NVIDIA Corporation \\
Technology & ORCL & Oracle Corporation \\
Technology & QCOM & QUALCOMM Incorporated \\
Technology & ROP & Roper Technologies, Inc. \\
Technology & SNPS & Synopsys, Inc. \\
Technology & STX & Seagate Technology Holdings plc \\
Technology & SWKS & Skyworks Solutions, Inc. \\
Technology & TXN & Texas Instruments Incorporated \\
Technology & VRSN & VeriSign, Inc. \\
Technology & WDC & Western Digital Corporation \\
Utilities & AEE & Ameren Corporation \\
Utilities & AEP & American Electric Power Company, Inc. \\
Utilities & AES & The AES Corporation \\
Utilities & CMS & CMS Energy Corporation \\
Utilities & CNP & CenterPoint Energy, Inc. \\
Utilities & D & Dominion Energy, Inc. \\
Utilities & DTE & DTE Energy Company \\
Utilities & DUK & Duke Energy Corporation \\
Utilities & ED & Consolidated Edison, Inc. \\
Utilities & EIX & Edison International \\
Utilities & ES & Eversource Energy \\
Utilities & ETR & Entergy Corporation \\
Utilities & EXC & Exelon Corporation \\
Utilities & FE & FirstEnergy Corp. \\
Utilities & LNT & Alliant Energy Corporation \\
Utilities & NEE & NextEra Energy, Inc. \\
Utilities & NI & NiSource Inc. \\
Utilities & NRG & NRG Energy, Inc. \\
Utilities & PEG & Public Service Enterprise Group Incorporated \\
Utilities & PNW & Pinnacle West Capital Corporation \\
Utilities & SO & The Southern Company \\
Utilities & SRE & Sempra \\
Utilities & WEC & WEC Energy Group, Inc. \\
Utilities & XEL & Xcel Energy Inc. \\
\bottomrule
\end{longtable}}

\ed